# The Twisted Spin Synchrotron

Ya.S. Derbenev

Randall Laboratory of Physics, University of Michigan
Ann Arbor, MI 48109-1120

**Abstract**

A "Figure 8" shaped synchrotron is suggested for use in accelerating polarized protons in the energy range below $\sim 20$ GeV.

The spin tune in such an accelerator does not ramp with energy and is equal to zero . Then the intrinsic spin resonances will not appear. A partial Siberian snake (solenoid) is proposed to be inserted into the ring in order to stabilize the spin against imperfect horizontal magnetic fields and implicate a small constant spin tune. The equilibrium polarization will then be horizontal (longitudinal in the solenoid).

The twisted spin synchrotron (TSS) can serve as a booster for high energy accelerators with polarized beams. Possibilities to use the TSS as a storage facility or a collider for proton or other particle beams with readily controlled beam polarization are also noted.

## 1 Introduction

When accelerating in a synchrotron, a polarized charged particle ( proton) beam encounters the depolarizing spin resonances associated with imperfect and focusing horizontal magnetic fields [1]. A spin resonance in an accelerator is similar to the well-known Nuclear Magnetic Resonance phenomenon. The difference is that the spin precession tune of a relativistic particle ramps with the energy $E = \gamma mc^2$:

$$\nu_{sp} = \gamma G,$$

where

$$G = (g-2)/2,$$

the magnetic moment anomaly is 1.79 for protons, and $-0.143$ for deuterons.

The non-controlled, incoherent spin flip practically kills the beam polarization after crossing a large number of resonances.

To overcome the spin resonances, special spin rotators, the so-called Siberian snakes, were proposed to be inserted into synchrotrons [1,2,3]. Totally controlled and coherent spin flip caused by a snake leads to self-cancellation of the spin precession phase advance in the vertical field (Fig. 1). This effect is similar to the *spin echo* phenomenon in NMR. The resulting spin tune then becomes independent of the particle energy, and crossing the spin resonances no longer occurs.

Snakes' designs are simple in the energy range above ~ 10 - 20 GeV. At lower energies, they become combersome (although not impossible to design for implication in proton rings).

The complex measures based on adiabatic or fast crossing of the spin resonances are proposed to preserve polarization in boosters and pre-accelerators [1,2,4,5].

In this paper, we will show a simple and efficient way to implicate the *spin echo* for accelerating polarized beams in the low energy range using especially the twisted cyclic accelerators.

## 2 Spin dynamics in a twisted synchrotron

Consider spin motion in a "Figure 8" shaped cyclic accelerator, i.e. a twisted ring (Fig. 2). The bending magnetic field $B_y$, and particle and spin precession angular velocities have opposite directions in the two loops of the accelerator. Assume at first a planar closed beam orbit. Then, according to the general equations of particle and spin motion [1], there is a simple ratio between the beam bend, $\Delta\theta$, and the spin precession phase advance, $\Delta\Psi$:

$$\Delta\vec{\Psi} = \gamma G \Delta\vec{\theta}, \quad \text{where} \quad d\vec{\theta} = \frac{\vec{v} \times d\vec{v}}{v^2}.$$

The resulting spin phase advance for a complete cycle is equal to zero:

$$\oint d\vec{\Psi} = \gamma G \oint d\vec{\theta} = \gamma(\vec{\theta_o} - \vec{\theta_o}) = 0.$$

The corresponding spin matrix is a unit, and the spin tune is equal to zero. This is a secular, i.e. resonance case: a horizontal magnetic field on the closed orbit will cause spin precession around a horizontal direction. In particular, the imperfection radial field $B_x(z)$ will cause a non-controlled spin flip, i.e. depolarization. To provide control and stability of the spin, insert a solenoid with field $B_s$ in a straight section. With no imperfections (i.e. assuming a unit spin matrix beyond the solenoid), the observed spin motion in the solenoid section will be a simple precession around the beam direction, with a phase advance per cycle of

$$\varphi_s = \frac{geB_s\ell}{2pc},$$

where $p^2c^2 = E^2 - m^2c^4$, and $\ell$ is the solenoid's length. According to a theorem on general properties of spin motion in a periodical field [1,6], the periodical polarization with the longitudinal direction in the solenoid section represents the periodical spin motion $\vec{n}(z)$, while the value $\nu_{sp} = \varphi_s/2\pi$ represents the tune of free spin precession around $\vec{n}$ ($\vec{n} \cdot \vec{s} = const$).

Polarization $\vec{n}$ is unique and stable if $\varphi_s \neq k$, where k is an integer, while the polarization transverse to $\vec{n}$ oscillates and disappears with time because of the spin tune dispersion. In practice to preserve polarization, a sufficient condition would be

$$\varepsilon << \nu_{sp} \qquad (1)$$

at $\nu_{sp} << 1$, where

$$\varepsilon = \gamma|\langle GB_x e^{i\Psi}\rangle|/\langle |B_y|\rangle,$$

and brackets $\langle...\rangle$ denote mean values around the beam orbit. The $\varepsilon$ value represents the resonance strength of the radial field for a closed orbit.

The solenoid can ramp with energy up to a maximum strength maintaining a constant $\nu_{sp}$ value.

The above-considered way of spin stabilization in a twisted accelerator is related in principle to the so-called partial Siberian snake method in conventional accelerators suggested earlier to stabilize the spin against the imperfection resonances during acceleration [2,5]. A successful test of the partial snake (solenoid) at AGS ($\nu_{sp} = 0.025$) showed the realization of condition (1) in the energy range $E \leq 20\,GeV$: no depolarization due to the imperfection resonances was observed [7][1].

However, the beam in the AGS experiments was still depolarized by the intrinsic resonances

$$\gamma G = k \pm \nu_y$$

($\nu_y$ is the vertical beam tune), since the spin tune, $\gamma G$, ramps with energy.

A principal property of a twisted synchrotron which makes it different from a conventional cyclic machine is that the spin tune, $\nu_{sp}$, does not ramp with energy. Therefore, the intrinsic resonances are eliminated similar to that in accelerators with full Siberian snakes.

Finally, the equilibrium polarization is horizontal (longitudinal in the solenoid section) in the TSS. If required, necessary spin matching for injected and ejected beams can be arranged by using appropriate spin rotators.

## 3  Possible uses of TSS

1. Generally, the twisted spin synchrotron can serve as a booster for succeeding proton accelerators with Siberian snakes. A horizontally (longitudinally) polarized beam can be injected into the TSS. With condition (1) satisfied, no additional spin cure is required to drive a polarized beam in TSS to an ejection energy (10 –20 GeV).

2. Implication of twist makes it simple and safe to accelerate and maintain polarized hadron beams (proton, deuteron, etc.) in storage facilities with internal targets.

---

[1]Earlier, this method was also successfully tested in the experiments with $e^{\pm}$ beams (Novosibirsk [8]) and low energy proton beam (IUCF [9]) to overcome a single imperfection resonance.

There are further possibilities to control the beam polarization in the target section by manipulating the solenoids (Fig. 3). In particular, longitudinal polarization of a deuteron beam can be easily obtained in the target section.

3. Finally, we note the possibility of a special "spinoff" (but with spin!): polarized colliding beams in a storage facility with a single beam track (Fig. 4).

# 4 Conclusion

The above-treated TSS concepts involve the *spin echo* principle in the low energy region, where use of full Siberian snakes is difficult or complicated. In the TSS, polarized beams become spin-transparent and simply spin-controlled. The possibilities of a polarized beam storage facility and a single track collider may be of interest meriting study in more detail.

# Acknowledgements

I acknowledge A.D. Krisch, J. Cameron, P. Schwandt, T. Roser, and A.S. Belov for their interest in this work and useful discussions.

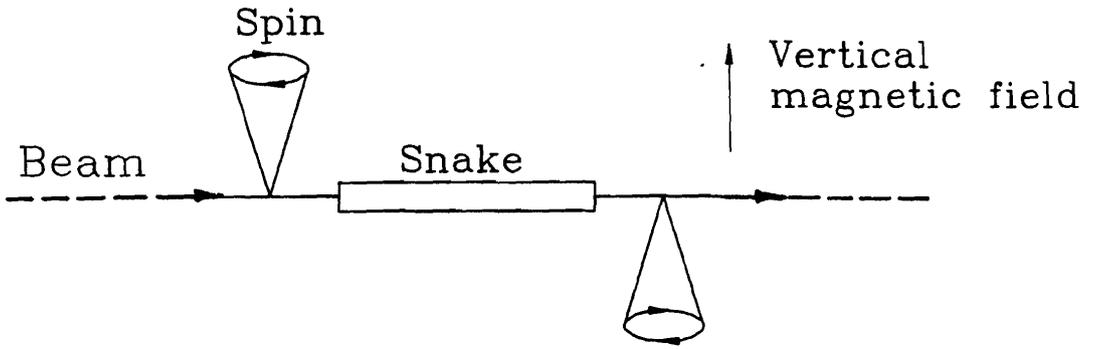

Fig.1. The Siberian snake's spin echo effect (side horizontal view). Spin flip caused by a snake results in a reverse of spin precession helicity.

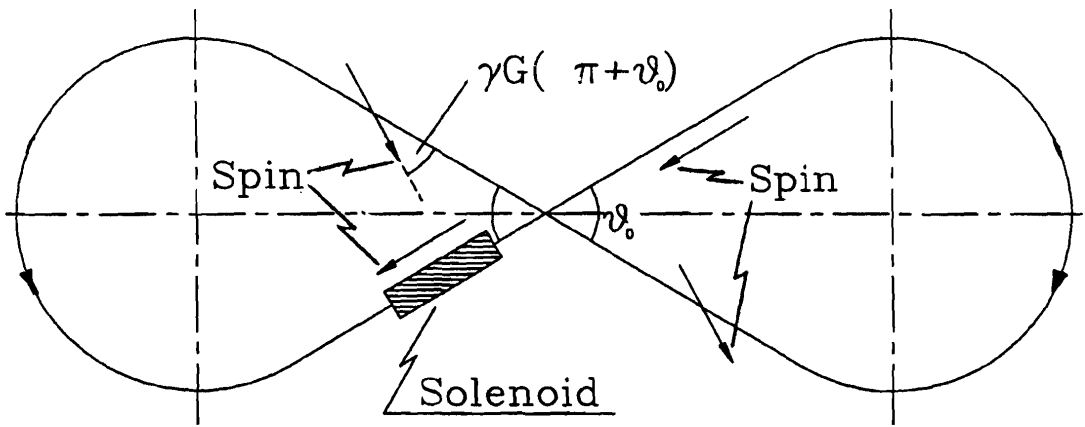

Fig. 2. Twisted spin synchrotron.

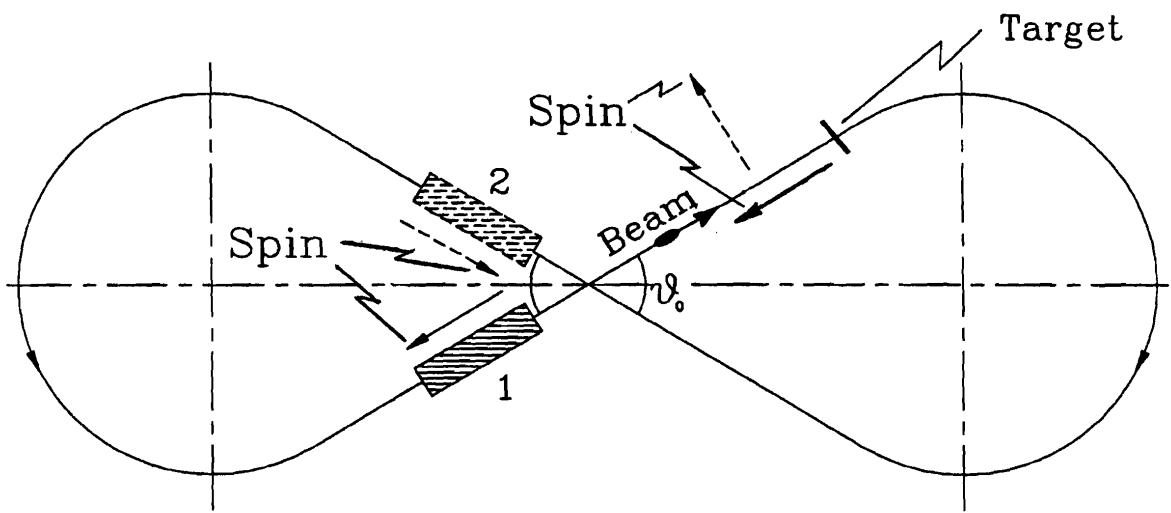

Fig. 3. Spin switching in TSS with an internal target.

Case 1 (thick line): longitudinal polarization, solenoid 2 off.
Case 2 (dashed line): transverse (horizontal) polarization
at $G\gamma(\pi + \vartheta_0) \approx (k + \frac{1}{2})\pi$, solenoid 1 off.

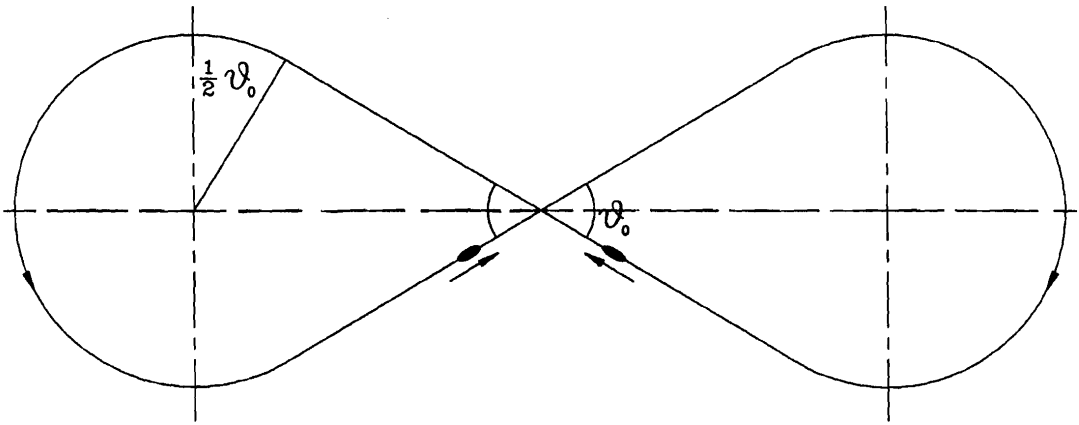

a) cross collisions

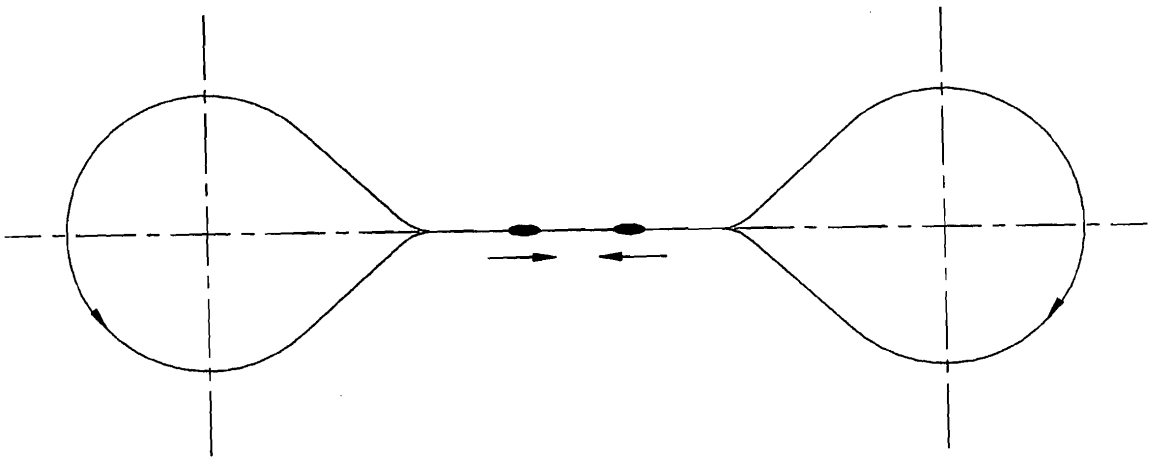

b) head-to-head collisions

Fig. 4. TSS as a collider